\RequirePackage{lineno}
\documentclass[prd,twocolumn,amsmath,amssymb]{revtex4}
\usepackage[colorlinks,linkcolor=blue,anchorcolor=blue,citecolor=blue,urlcolor=blue]{hyperref}
\usepackage{graphicx}
\usepackage{dcolumn}
\usepackage{bm}
\usepackage{overpic}
\usepackage{rotating}
\usepackage{epstopdf}
\usepackage{color}
\newcommand{\BR}{{\cal B}}

\newcommand{\ee}{e^+e^-}
\newcommand{\EE}{e^+e^-}

\newcommand{\pip}{\pi^+}
\newcommand{\pim}{\pi^-}

\newcommand{\pp}{\pi^+\pi^-}
\newcommand{\PP}{\pi^+\pi^-}
\newcommand{\KK}{K^+K^-}

\newcommand{\unitegev}{~{\rm GeV}}
\newcommand{\unitemev}{~{\rm MeV}}

\newcommand{\unitmgev}{~{\rm GeV}/c^2}
\newcommand{\unitmmev}{~{\rm MeV}/c^2}

\newcommand{\process}{$\EE\to\phi\pip\pim~$}

\def\babar{\mbox{\slshape B\kern-0.1em{\small A}\kern-0.1em B\kern-0.1em{\small A\kern-0.2em R}}}
\begin{document}
\normalsize
\parskip=5pt plus 1pt minus 1pt
\title{Measurement of \process cross sections at center-of-mass energies from 2.00 to 3.08 GeV}
\author{
  M.~Ablikim$^{1}$, M.~N.~Achasov$^{13,b}$, P.~Adlarson$^{73}$, R.~Aliberti$^{34}$, A.~Amoroso$^{72A,72C}$, M.~R.~An$^{38}$, Q.~An$^{69,56}$, Y.~Bai$^{55}$, O.~Bakina$^{35}$, I.~Balossino$^{29A}$
% ...
  \\
  \vspace{0.2cm}
  (BESIII Collaboration)\\
  \vspace{0.2cm} {\it
    $^{1}$ Institute of High Energy Physics, Beijing 100049, People's Republic of China\\
% ...
$^{13}$ G.I. Budker Institute of Nuclear Physics SB RAS (BINP), Novosibirsk 630090, Russia\\
% ...
$^{29}$ INFN Sezione di Ferrara, (A)INFN Sezione di Ferrara, I-44122, Ferrara, Italy; (B)University of Ferrara, I-44122, Ferrara, Italy\\
% ...
$^{34}$ Johannes Gutenberg University of Mainz, Johann-Joachim-Becher-Weg 45, D-55099 Mainz, Germany\\
    $^{35}$ Joint Institute for Nuclear Research, 141980 Dubna, Moscow region, Russia\\
% ...
$^{38}$ Liaoning Normal University, Dalian 116029, People's Republic of China\\
% ...
$^{55}$ Southeast University, Nanjing 211100, People's Republic of China\\
    $^{56}$ State Key Laboratory of Particle Detection and Electronics, Beijing 100049, Hefei 230026, People's Republic of China\\
% ...
$^{69}$ University of Science and Technology of China, Hefei 230026, People's Republic of China\\
% ...
$^{72}$ University of Turin and INFN, (A)University of Turin, I-10125, Turin, Italy; (B)University of Eastern Piedmont, I-15121, Alessandria, Italy; (C)INFN, I-10125, Turin, Italy\\
    $^{73}$ Uppsala University, Box 516, SE-75120 Uppsala, Sweden\\
% ...
$^{b}$ Also at the Novosibirsk State University, Novosibirsk, 630090, Russia\\
% ...
  }
}
\date{\today}
\begin{abstract}
  {
    Using data corresponding to an integrated luminosity of $651~\mathrm{pb}^{-1}$ accumulated at 22 center-of-mass energies from 2.00 to 3.08 GeV by the BESIII experiment, the process \process is studied. The cross sections for \process are consistent with previous results, but with improved precision. To measure the mass and width of the structure observed in the cross section line shape, a combine fit is performed after enhancing the contribution from $\phi f_{0}(980)$. The fit reveals a structure with the mass of $M=2178\pm20\pm5\unitmmev$ and the width of $\varGamma=140\pm36\pm16\unitemev$, where the first uncertainties are statistical and the second ones are systematic.
  }
\end{abstract}
\maketitle
\section{\boldmath Introduction}
{
The study of the hadron spectrum is important to understand the non-perturbative behavior of quantum chromodynamics (QCD). For the low-energy region, the vector mesons $\rho$, $\omega$, $\phi$ and their low-lying excited states are copiously produced in $e^{+}e^{-}$ collision experiments. The experimental results for these states have been tabulated by the Particle Data Group (PDG)~\cite{PDG}, but the higher lying excitations are not fully identified yet, especially in the region around 2.0 GeV. Further measurements are needed to resolve the situation involving resonances such as the $\rho(2000)$, $\rho(2150)$ and $\phi(2170)$ states.

The $\phi(2170)$ resonance was first observed by the \babar\ Collaboration via the initial state radiation (ISR) process $e^{+}e^{-} \to \gamma_{\rm ISR}\phi f_{0}(980)$~\cite{observed1, observed2}, and later confirmed by the Belle, BESII, and BESIII experiments~\cite{confirmed:1, confirmed:2, renew, update:1, update:2}. This observation stimulated speculation that the $\phi(2170)$ resonance might be a strangeonium counterpart of the charmonium resonance $\psi(4260)$ due to similarities in their production and decay pattern~\cite{speculation}. Considerable efforts have been made theoretically to understand the nature of the $\phi(2170)$ resonance and abundant interpretations have been proposed, including a traditional $s \bar s$ state~\cite{ssbar:1, ssbar:2, ssbar:3, ssbar:4, ssbar:5, ssbar:6}, an $s\bar s g$ hybrid~\cite{speculation, ssbarg}, an $ss \bar s \bar s$ tetra-quark state~\cite{4quark:1,4quark:2,4quark:3,4quark:4,4quark:5,4quark:6,4quark:7,4quark:8}, a $\Lambda \bar \Lambda$ bound state~\cite{llbar:1,llbar:2,llbar:3,llbar:4,llbar:5} and an ordinary $\phi K \bar K$ or $\phi f_{0}(980)$ resonance produced by interactions between the final state particles~\cite{ordinary:1,ordinary:2}. The model predictions differ in both mass and width of the resonance. Further experimental studies are therefore crucial to clarify its nature.

Though many experiments have been carried out to study the $\phi(2170)$ resonance~\cite{observed1, observed2, confirmed:1, confirmed:2, renew, update:1, update:2, KKeta}, the results of the measurements vary substantially. For example, the mass and width of the $\phi(2170)$ resonance obtained from the process $e^{+}e^{-}\to\gamma_{\rm ISR}\phi\pi^{+}\pi^{-}$~\cite{confirmed:2} shows smaller values than other experimental measurements. Recently, more studies related to the $\phi(2170)$ resonance have been carried out by the BESIII experiment. A partial wave analysis {of} the $e^{+}e^{-}\to K^{+}K^{-} \pi^{0}\pi^{0}$ process~\cite{KKPi0Pi0} found that the partial widths of the $\phi(2170)$ resonance are sizable for the $K(1460)^{+}K^{-}$, $K_{1}(1400)^{+}K^{-}$, and $K_{1}(1270)^{+}K^{-}$ decay channels, but much smaller for $K^{*}(892)^{+}K^{*}(892)^{-}$ and $K^{*}(1410)^{+}K^{-}$. Several theoretical expectations are challenged by the results according to Ref.~\cite{ssbar:1}. Attempts have also been made to study channels with simpler topologies, including {the processes} $e^{+}e^{-}\to K^{+}K^{-}$, where a resonance with a mass of $(2239.2 \pm 7.1 \pm 11.3)$ MeV/$c^{2}$ and a width of $(139.8 \pm 12.3 \pm 20.6)$ MeV is seen~\cite{KK}; $e^{+}e^{-} \to \phi K^{+}K^{-}$~{\cite{phiKK}}, where a sharp enhancement is observed in the Born cross section line-shape at a {center-of-mass~(c.m.) energy} of $\sqrt{s}$ = 2.2324 GeV; $e^{+}e^{-} \to \phi\eta'$~\cite{phietaprime}, where a resonance with a mass of $(2177.5 \pm 5.1 \pm 18.6)$ MeV/$c^{2}$ and a width of $(149.0 \pm 15.6 \pm 8.9)$ MeV is seen; $e^{+}e^{-} \to \omega\eta$~\cite{omegaeta}, a resonance with a mass of $(2179 \pm 21 \pm 3)$ MeV/$c^{2}$ and a width of $(89 \pm 28 \pm 5)$ MeV is observed with a significance of 6.1$\sigma$; $e^{+}e^{-} \to \phi\eta$~\cite{phieta}, a resonant structure is observed with parameters determined to be M = ($2163.5 \pm 6.2 \pm 3.0$)~MeV/$c^{2}$ and $\varGamma$ = ($31.1^{+21.1}_{-11.6}$ $\pm$ 1.1)~MeV; and $e^{+}e^{-} \to K_{S}^{0}K_{L}^{0}$~\cite{KSKL}, a resonant structure around 2.2 GeV is observed, with a mass and width of $2273.7 \pm 5.7 \pm 19.3$~MeV/$c^{2}$ and $86 \pm 44 \pm 51$ MeV respectively. The Breit-Wigner parameters of $\phi(2170)$ are not consistent between the different studies, especially concerning the width.

In addition, a resonance-like structure which we called R(2400) might exist around 2.4 GeV in the $\phi\pip\pim$ cross section line-shape. The R(2400) was first studied by the Belle~\cite{confirmed:2} experiment. Later, Shen and Yuan~\cite{fitcombine} performed a fit to the R(2400) structure using the combined data of the Belle and \babar\ experiments. The mass and the width are determined to be (2436~$\pm$~26)~MeV/$c^{2}$ and (121~$\pm$~35)~MeV, respectively. However, its statistical significance is less than $3\sigma$. An interpretation is proposed for R(2400) as a partner state of the $\phi(2170)$ resonance~\cite{X2400}. Therefore, a precise measurement of \process is desirable to establish the mass and width of the $\phi(2170)$ resonance and to search for the possible structure near 2.4~GeV.

In this paper, the measurement of cross sections for the process \process at 22 center-of-mass energies ($\sqrt{s}$) is reported from 2.00 to 3.08 GeV.
}
\section{\boldmath Detector and data samples}
{
The BESIII detector~\cite{Ablikim:2009aa} records symmetric $e^+e^-$ collisions provided by the BEPCII storage ring~\cite{Yu:IPAC2016-TUYA01}, which operates with a peak luminosity of $1\times10^{33}$~cm$^{-2}$s$^{-1}$ in the {c.m. energy} range between 2.0000 and 4.9000~GeV. BESIII has collected large data samples in this energy region~\cite{Ablikim:2019hff}. The cylindrical core of the BESIII detector covers 93\% of the full solid angle and consists of a helium-based multilayer drift chamber~(MDC), a plastic scintillator time-of-flight system~(TOF), and a CsI(Tl) electromagnetic calorimeter~(EMC), which are all enclosed in a superconducting solenoidal magnet providing a 1.0~T magnetic field. The solenoid is supported by an octagonal flux-return yoke with resistive plate counter muon identification modules interleaved with steel. The charged-particle momentum resolution at $1~{\rm GeV}/c$ is $0.5\%$, and the d$E$/d$x$ resolution is $6\%$ for electrons from Bhabha scattering. The EMC measures photon energies with a resolution of $2.5\%$ ($5\%$) at $1$~GeV in the barrel (end-cap) region. The time resolution in the TOF barrel region is 68~ps, while that in the end-cap region is 110~ps. The end-cap TOF system was upgraded in 2015 using multi-gap resistive plate chamber technology, providing a time resolution of 60~ps~\cite{etof1,etof2,etof3}.

Simulated Monte Carlo (MC) samples of signal and background processes are produced to optimize the event selection criteria, determine the detection efficiency and estimate the background contamination. The response of the detector is reproduced using a {\sc geant4}-based~\cite{GEANT4} {MC} simulation software package, which includes the geometric and material description of the BESIII detector, the detector response and digitization models.

Background samples of QED processes are produced with the {\sc babayaga}~\cite{BABAYAGA} generator and inclusive hadronic processes are generated with the {\sc luarlw}~\cite{luarlw} generator.

The signal MC samples of the process $e^+ e^- \to \phi \pi^+ \pi^-$ are generated from a uniform distribution in phase space (PHSP) reweighted by an amplitude analysis. We simulate one million events at each energy point. The signal MC samples are used to determine the reconstruction efficiency, and the correction factors for ISR and vacuum polarization (VP).
}
\section{\boldmath Event Selection and Background Analysis}
{
Signal events of the $e^{+}e^{-} \to \phi \pi^{+} \pi^{-}$ process are reconstructed via the $\phi \to K^{+} K^{-}$ decay. Charged track candidates are reconstructed from hits in the MDC and need to satisfy $|\cos\theta|<0.93$, where $\theta$ is the polar angle with respect to the symmetry axis of the MDC. The closest approach to the interaction point is required to be less than 10 cm along the symmetry axis and less than 1 cm in the perpendicular plane. Combined TOF and d$E$/d$x$ information is used to perform the particle identification~(PID), obtaining probabilities for the $\pi, K$ and $p$ hypotheses. The particle type with the largest probability is assigned to each track. Since the tracking efficiency decreases sharply in the low momentum region below 0.5~GeV/$c$, and most kaon candidates are expected to have a low momentum, one kaon is allowed to be missing in this study to increase the selection efficiency. Including events with one missing kaon increases number of signal events by factor of 3 at 2.00 GeV and 30\% at 3.08 GeV. A candidate event is, therefore, expected to have two pions and at least one kaon reconstructed.

A vertex fit to the $\pi^{+}\pi^{-}K^{\pm}$ combination is then applied and required to have converged for an event to be kept for further analysis. For events with four charged tracks, both $\pi^{+}\pi^{-}K^{+}$ and $\pi^{+}\pi^{-}K^{-}$ combinations are tested. {Under the hypothesis that one kaon is missing}, a one-constraint (1C) kinematic fit is performed to the combinations that {are kept after the} vertex fit. For each event, the $\pi^{+}\pi^{-}K^{\pm}$ combination with the smallest $\chi^{2}$ of the 1C kinematic fit ($\chi^{2}_{\rm1C}$($\pi^{+}\pi^{-}KK_{miss}$)) is retained. Finally, events with $\chi^{2}_{\rm 1C} \geq 10$ are rejected. After applying the selection criteria, we use the momenta of the particles obtained from the kinematic fit in the further analysis.

Events passing the selection criteria described above are shown in Fig.~\ref{figure:mass} for the data at $\sqrt{s} = 2.1250$~GeV. The invariant mass of the $K^{+}K^{-}$ {pairs} shows a clear signal band around the $\phi$ mass. The enhancement around 0.98~GeV/$c^{2}$ in the $\pip\pim$ invariant mass indicates a correlation between $f_{0}(980)$ and $\phi$ production due to the process $e^{+}e^{-} \to \phi f_{0}(980)$.

\begin{figure}[!ht]
  \begin{center}
    \begin{overpic}[width=0.47\textwidth]{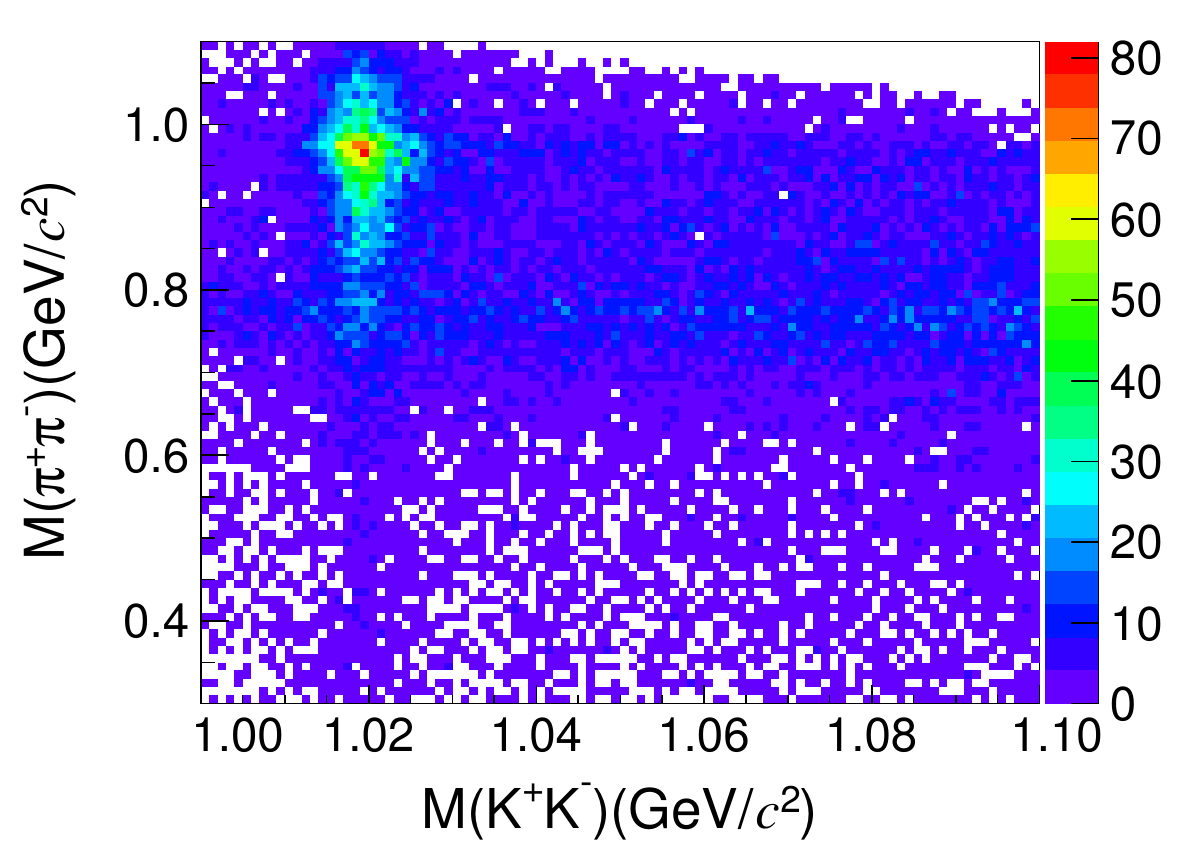}
    \end{overpic}
  \end{center}
  \caption{Distribution of $M_{\PP}$ versus $M_{\KK}$ for the data at $\sqrt{s}=2.1250$~GeV.}
  \label{figure:mass}
\end{figure}

The distribution of {the} $\KK$ invariant mass is shown in Fig.~\ref{figure:signal}. The range of $|M_{K^{+}K^{-}} - m_{\phi}| < 0.01$~GeV/$c^{2}$ is regarded as the signal region in the following study, where $m_{\phi}$ = 1019.461~MeV/$c^{2}$ is the world average $\phi$ mass from the PDG~\cite{PDG}. The sideband regions, defined as [0.995,1.005] and [1.035,1.045]~GeV/$c^{2}$, are used to study non-$\phi$ background contributions.

\begin{figure}[!ht]
  \begin{center}
    \begin{overpic}[width=8.5cm,height=6.5cm,angle=0]{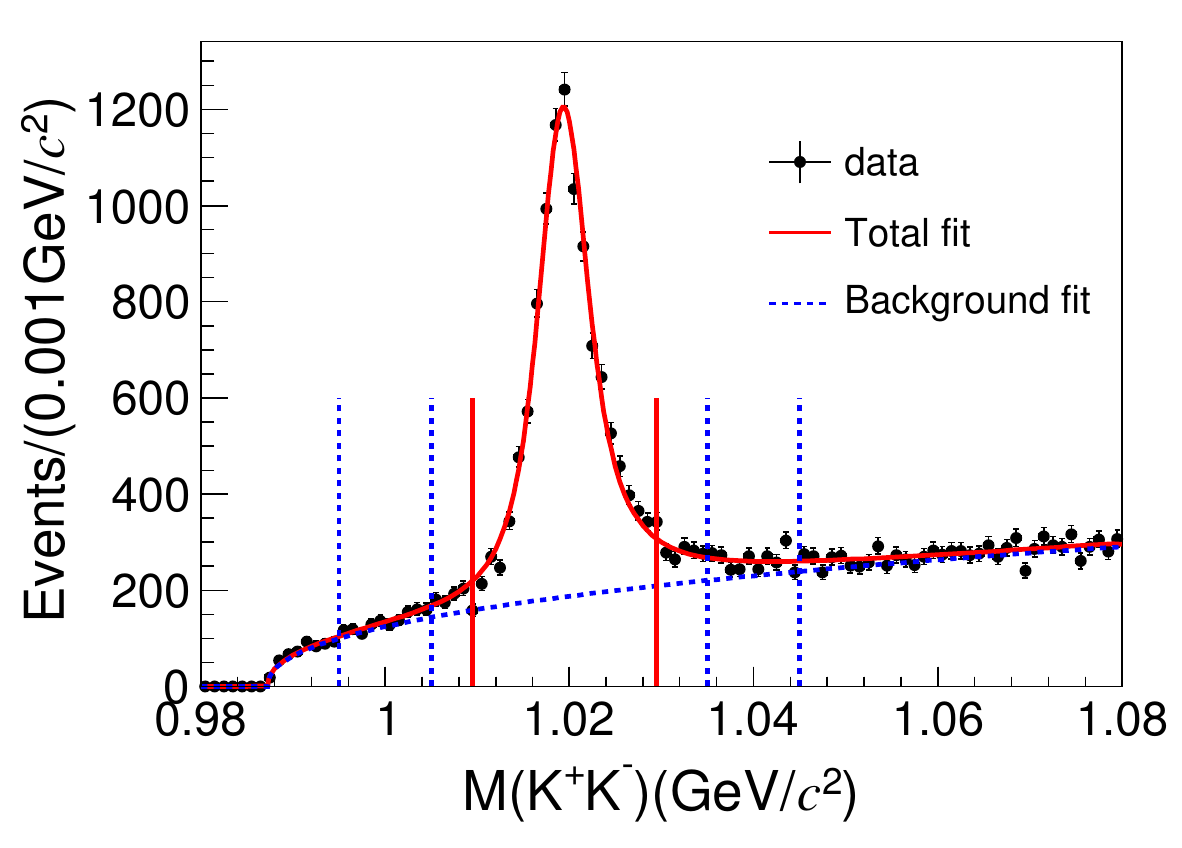}
    \end{overpic}
  \end{center}
  \caption{Fit to the $M_{K^{+}K^{-}}$ distribution for the data at $\sqrt{s}=2.1250$~GeV: the signal is described by a P-wave Breit-Wigner (BW) function convolved with a Gaussian function, and the background is described by a reversed ARGUS function. The range between the two red vertical solid lines is regarded as signal region, and the ranges between the two blue vertical dashed lines on each side of the signal peak are regarded as the sideband regions.}
  \label{figure:signal}
\end{figure}

An accumulation of events exists around the mass of the $\rho$ meson in Fig~\ref{figure:mass}. This indicates a non-negligible background contribution from the $e^{+}e^{-} \to \rho K^+ K^-$ process. Based on a study of the $\phi$ sideband and an analysis of the inclusive MC sample, the $\EE \to K^{*}(892)K^{\pm}\pi^{\pm}$ process is found to be the dominant background source. Peaking background in the $\phi$ signal region is negligible.
}
\section{\boldmath Signal Yields}
{
The $e^{+}e^{-} \to \phi\pi^{+}\pi^{-}$ signal yields are obtained from unbinned maximum likelihood fits to the $\KK$ invariant mass in the region [$2\cdot m_{K^\pm}$, 1.08]~GeV/$c^{2}$. {In the fit,} the $\phi$ peak is modeled as a P-wave BW function convolved with a Gaussian function to account for a difference in detector resolution and an offset in calibration between data and the MC simulation~\cite{phiKK}. The P-wave BW function is defined in the form
\begin{center}
  \begin{equation}
    f(m) = |A(m)|^{2} \cdot p^{2l+1},
  \end{equation}
  \begin{equation}
    A(m) = \frac{1}{m^{2} - m_{\phi}^{2} + im\varGamma(m)} \frac{B(p)}{B(p')} ,
  \end{equation}
  \begin{equation}
    \varGamma(m) = {\left( \frac{p}{p'} \right)^{2l+1} \left( \frac{m_{\phi}}{m} \right) \varGamma_{\phi} \left[ \frac{B(p)}{B(p')} \right]} ,
  \end{equation}
  \begin{equation}
    B(p) = \frac{1}{\sqrt{1+(Rp)^{2}}} ,
  \end{equation}
\end{center}
where $p$ is the momentum of the kaon in the rest frame of the $K^{+}K^{-}$ system, $p'$ is the momentum of the kaon at the $\phi$ peak mass, and $\varGamma_{\phi}$ is the width of the $\phi$ resonance~\cite{PDG}. The angular momentum ($l$) is equal to one, $B(p)$ is the Blatt-Weisskopf form factor, and $R = 3$~GeV$^{-1}$ is the radius of the centrifugal barrier~\cite{phifit}.

Since no peaking background is expected in the signal area, the background is parameterized with a reversed ARGUS function~\cite{Argus}. The parameters of the Gaussian function and the reversed ARGUS function are determined in a fit to the data. The fit result at $\sqrt{s} = 2.1250$~GeV is shown in Fig.~\ref{figure:signal}. We obtain a similar fit quality for all center-of-mass energies.
}
\section{\boldmath Cross section calculation}
{
\label{section:crosssection}
The dressed cross section of the process \process is calculated by:
\begin{center}
  \begin{equation}
    \sigma=\frac{N^{\rm obs}}{\mathcal{L} \cdot (1+\delta^{\gamma}) \cdot \epsilon \cdot \cal{B}} ,
    \label{equation:crosssection}
  \end{equation}
\end{center}
where $N^{\rm obs}$ is the signal yield; $\mathcal{L}$ is the integrated luminosity; $(1+\delta^{\gamma})$ is the ISR correction factor; $\epsilon$ is the detection efficiency and $\cal B$ is the branching fraction of the decay $\phi\to K^{+}K^{-}$. The ISR correction factor is handled by generator {\sc conexc}~\cite{CONEXC}, depending on the input cross sections. The lowest-order Born cross section is defined as $\sigma^{\rm B}=\sigma^{\rm D} / (1/|1 - \Pi|^{2})$, where $1/|1 - \Pi|^{2}$ is the VP correction factor, available from F. Jegerlehner group~\cite{VPfactor}.

To adequately describe the data in our MC simulation, the signal MC is generated from a uniform distribution in PHSP reweighted by an amplitude analysis. The quasi-two-body decay amplitudes in the sequential decays are constructed using covariant tensor amplitudes~\cite{Likelihood_Amplitudes}. The \process process is found to be well described by four subprocesses: $e^{+}e^{-} \to \phi f_{0}(980), ~\phi \sigma, ~\phi f_{0}(1370)$ and $\phi f_{2}(1270)$. The intermediate states are parametrized with relativistic BW functions, except for the $\sigma$ and $f_{0}(980)$, which are described with using the model described in \cite{sigmapole} and by a Flatt$\acute{e}$ formula~\cite{Flatte}, respectively. The resonance parameters of the $f_{0}(980)$ and the wide resonance $\sigma$ in the fit are fixed to those in Ref.~\cite{Flatte} and Refs.~\cite{Flatte,sigmapole}, respectively, and those of other intermediate states are fixed to the PDG values. The relative magnitudes and phases of the individual intermediate processes are determined by performing an unbinned maximum likelihood fit using MINUIT~\cite{MINUIT}. To describe the background below the $\phi$ peak, sideband events are added to the likelihood with negative weights. For a few low-statistic points, the fitted parameters obtained at the most adjacent high-statistic energies are applied.

The reweighted signal MC simulation has reasonable agreement with the experimental data at all center-of-mass energies. The comparison of the MC simulation and experimental data in the signal region for the $M(\pp)$ distribution at $\sqrt{s}=2.1250\unitegev$ is shown in Fig.~\ref{figure:compare}.

\begin{figure}[!ht]
  \begin{center}
    \begin{overpic}[width=0.5\textwidth]{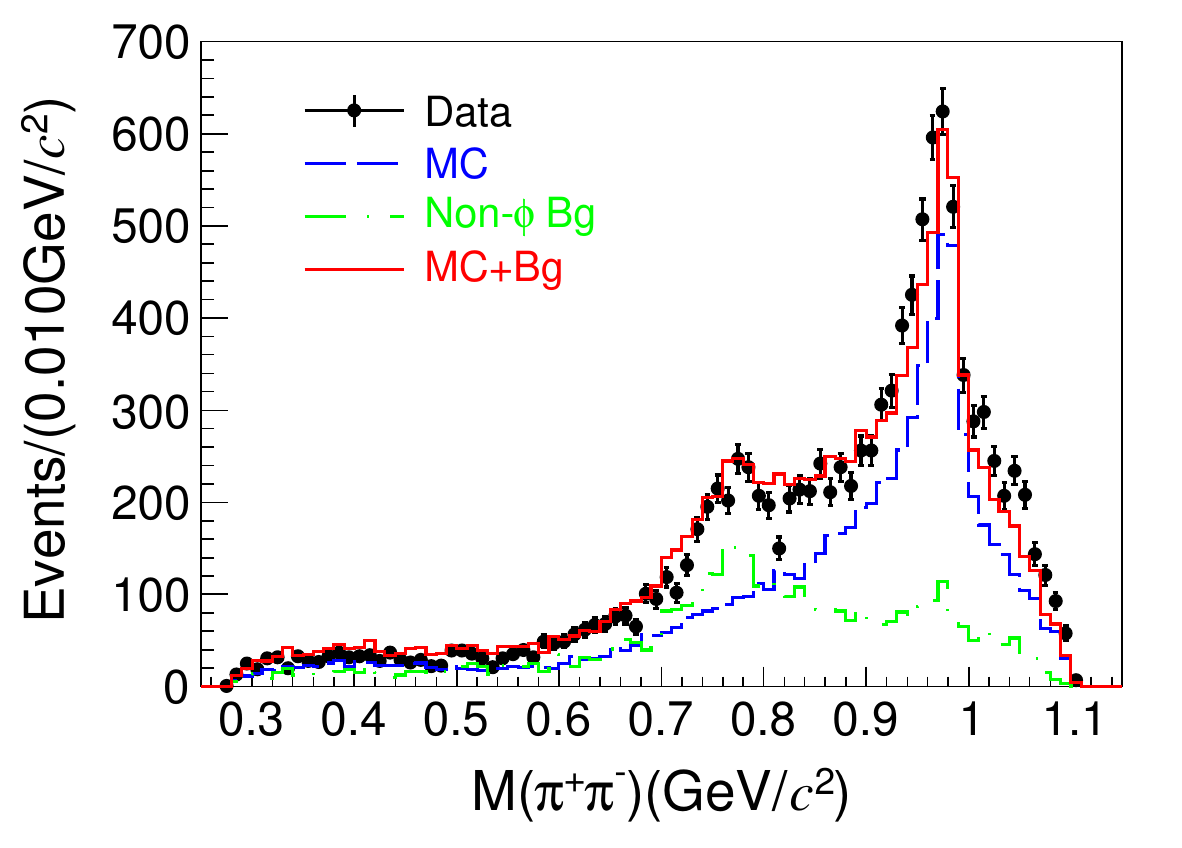}
    \end{overpic}
  \end{center}
  \caption{The invariant mass distribution of the $\pi^{+}\pi^{-}$ candidates for the data at $\sqrt{s}=2.1250$~GeV. The black dots with uncertainties are experimental data, the blue dashed line is the reweighted signal MC distribution, the green dash-dotted line is the non-$\phi$ background estimated from the $\phi$ sideband region and the red solid line is the sum of the former two distributions.}
  \label{figure:compare}
\end{figure}

The efficiency $\epsilon$ and the ISR correction factor ($1+\delta^{\gamma}$) depend on the input cross section line-shape and need to be determined using an iterative procedure. The \babar\ result~\cite{renew} is used as the initial input cross section and the updated cross section is obtained through the resulting MC simulation. This procedure is repeated until the measured cross section converges. Dressed cross sections for \process at each energy point are listed in Table~\ref{table:crosssection}, together with Born cross sections and VP correction factors. The measured dressed cross sections are shown in Fig.~\ref{figure:phipp}.

\begin{figure}[!ht]
  \begin{center}
    \begin{overpic}[width=0.5\textwidth]{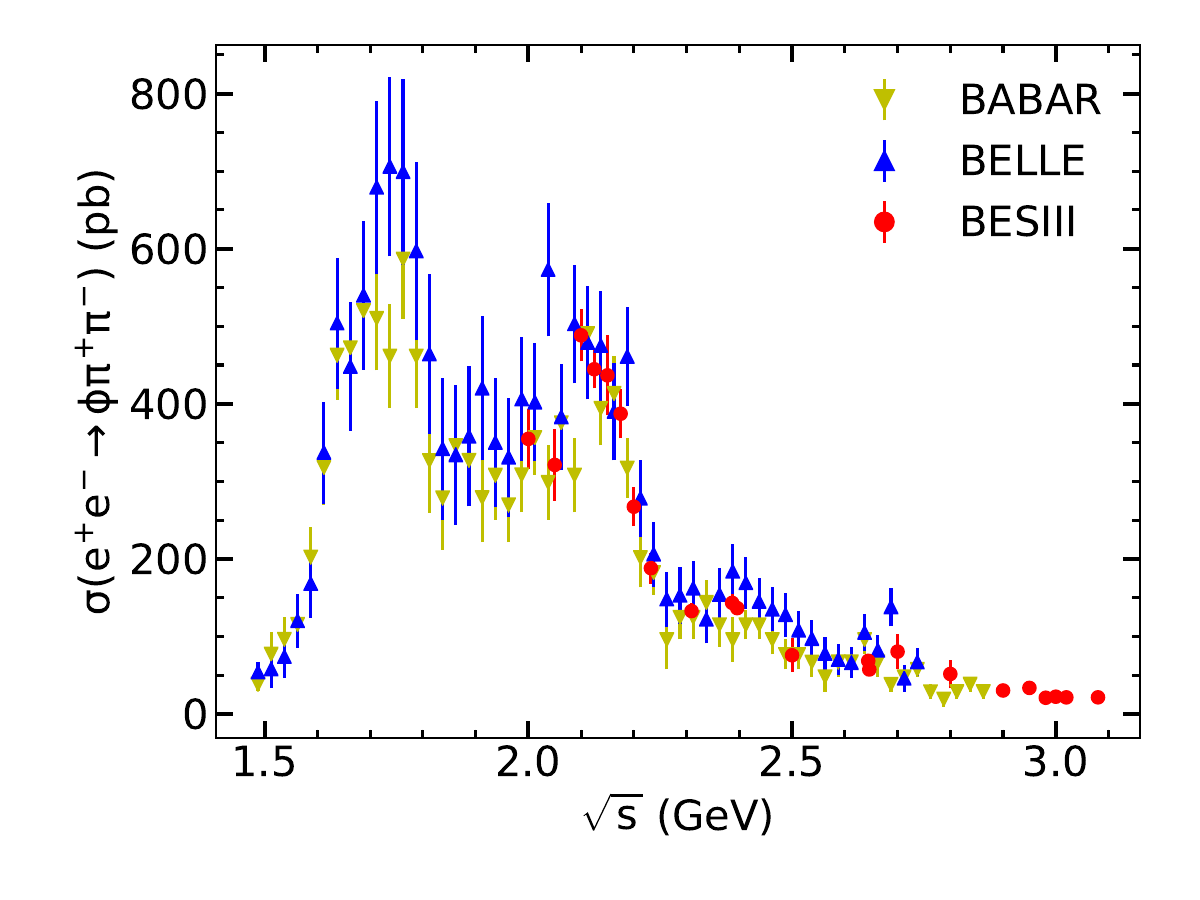}
    \end{overpic}
  \end{center}
  \caption{The measured dressed cross section of the process \process. Red dots with errorbars represent the measurement from BESIII. Yellow downside triangles and Blue upside triangles represent the measurements from \babar\ and Belle, respectively. The uncertainties incorporate statistical and uncorrelated systematic uncertainties.}
  \label{figure:phipp}
\end{figure}

\begin{table*}[!ht]
  \centering
  \caption{\small Cross sections of the process \process at different center-of-mass energies. The $\mathcal{L}$ is the integrated luminosity; $N^{\rm obs}$ is the yield of signal events; $1+\delta^{\gamma}$ is the ISR correction factor; $1/[1 - \Pi]^{2}$ is the VP correction factor; $\epsilon$ is the detection efficiency; $\sigma^{\rm B}$ is the Born cross section and $\sigma^{\rm D}$ is the dressed cross section. The uncertainties are statistical, uncorrelated systematic and correlated systematic uncertainties, respectively.}
  \label{table:crosssection}
  \resizebox{1.9\columnwidth}{!}
  {
    \begin{tabular}{c|c|rcl|c|c|c|rrrclll|rrrclll}
      \hline\hline
      $\sqrt{s}$ (GeV) & $\mathcal{L}$ (pb$^{-1}$) & \multicolumn{3}{c|}{$N^{\rm obs}$} & ($1+\delta^{\gamma}$) & $1/|1 - \Pi|^{2}$ & $\epsilon$ & \multicolumn{7}{c|}{$\sigma^{\rm B}$(pb)} & \multicolumn{7}{c}{$\sigma^{\rm D}$(pb)}                                                                                                           \\
      \hline
      2.0000           & 10.1                      & 577                                & $\pm$                 & 46                & 0.98       & 1.04                                      & 0.34                                     & 342.1 & $\pm$ & 27.7 & $\pm$ & 19.3 & $\pm$ & 15.3 & 354.7 & $\pm$ & 28.8 & $\pm$ & 20.1 & $\pm$ & 15.9 \\
      2.0500           & 3.34                      & 191                                & $\pm$                 & 24                & 0.96       & 1.04                                      & 0.38                                     & 309.4 & $\pm$ & 39.7 & $\pm$ & 14.9 & $\pm$ & 13.8 & 321.2 & $\pm$ & 41.3 & $\pm$ & 15.5 & $\pm$ & 14.4 \\
      2.1000           & 12.2                      & 1100                               & $\pm$                 & 51                & 0.95       & 1.04                                      & 0.40                                     & 470.0 & $\pm$ & 21.8 & $\pm$ & 11.8 & $\pm$ & 21.0 & 488.3 & $\pm$ & 22.7 & $\pm$ & 12.2 & $\pm$ & 21.8 \\
      2.1250           & 108                       & 9372                               & $\pm$                 & 144               & 0.95       & 1.04                                      & 0.42                                     & 427.9 & $\pm$ & 6.6  & $\pm$ & 12.7 & $\pm$ & 19.1 & 444.6 & $\pm$ & 6.8  & $\pm$ & 13.2 & $\pm$ & 19.9 \\
      2.1500           & 2.84                      & 220                                & $\pm$                 & 20                & 0.96       & 1.04                                      & 0.37                                     & 419.9 & $\pm$ & 38.2 & $\pm$ & 25.6 & $\pm$ & 18.8 & 436.7 & $\pm$ & 39.7 & $\pm$ & 26.6 & $\pm$ & 19.5 \\
      2.1750           & 10.6                      & 760                                & $\pm$                 & 39                & 1.00       & 1.04                                      & 0.37                                     & 372.3 & $\pm$ & 19.1 & $\pm$ & 17.2 & $\pm$ & 16.7 & 387.2 & $\pm$ & 19.9 & $\pm$ & 17.9 & $\pm$ & 17.3 \\
      2.2000           & 13.7                      & 706                                & $\pm$                 & 38                & 1.08       & 1.04                                      & 0.37                                     & 257.0 & $\pm$ & 13.9 & $\pm$ & 15.9 & $\pm$ & 11.5 & 267.2 & $\pm$ & 14.5 & $\pm$ & 16.5 & $\pm$ & 12.0 \\
      2.2324           & 11.9                      & 435                                & $\pm$                 & 29                & 1.19       & 1.04                                      & 0.33                                     & 180.6 & $\pm$ & 12.2 & $\pm$ & 12.4 & $\pm$ & 8.1  & 188.0 & $\pm$ & 12.7 & $\pm$ & 12.9 & $\pm$ & 8.4  \\
      2.3094           & 21.1                      & 587                                & $\pm$                 & 37                & 1.19       & 1.04                                      & 0.36                                     & 127.5 & $\pm$ & 8.2  & $\pm$ & 3.3  & $\pm$ & 5.7  & 132.7 & $\pm$ & 8.5  & $\pm$ & 3.4  & $\pm$ & 5.9  \\
      2.3864           & 22.5                      & 697                                & $\pm$                 & 37                & 1.13       & 1.04                                      & 0.39                                     & 137.6 & $\pm$ & 7.3  & $\pm$ & 5.4  & $\pm$ & 6.2  & 143.3 & $\pm$ & 7.6  & $\pm$ & 5.7  & $\pm$ & 6.4  \\
      2.3960           & 66.9                      & 1977                               & $\pm$                 & 65                & 1.13       & 1.04                                      & 0.39                                     & 131.1 & $\pm$ & 4.3  & $\pm$ & 5.1  & $\pm$ & 5.9  & 136.5 & $\pm$ & 4.5  & $\pm$ & 5.3  & $\pm$ & 6.1  \\
      2.5000           & 1.10                      & 18                                 & $\pm$                 & 5                 & 1.21       & 1.04                                      & 0.38                                     & 72.8  & $\pm$ & 20.2 & $\pm$ & 3.5  & $\pm$ & 3.3  & 75.7  & $\pm$ & 21.1 & $\pm$ & 3.7  & $\pm$ & 3.4  \\
      2.6444           & 33.7                      & 501                                & $\pm$                 & 33                & 1.31       & 1.04                                      & 0.34                                     & 65.9  & $\pm$ & 4.4  & $\pm$ & 3.8  & $\pm$ & 2.9  & 68.5  & $\pm$ & 4.5  & $\pm$ & 3.9  & $\pm$ & 3.1  \\
      2.6464           & 34.0                      & 423                                & $\pm$                 & 29                & 1.31       & 1.04                                      & 0.34                                     & 55.3  & $\pm$ & 3.9  & $\pm$ & 3.1  & $\pm$ & 2.5  & 57.4  & $\pm$ & 4.0  & $\pm$ & 3.3  & $\pm$ & 2.6  \\
      2.7000           & 1.03                      & 22                                 & $\pm$                 & 6                 & 1.64       & 1.04                                      & 0.33                                     & 77.4  & $\pm$ & 21.5 & $\pm$ & 4.2  & $\pm$ & 3.5  & 80.4  & $\pm$ & 22.3 & $\pm$ & 4.4  & $\pm$ & 3.6  \\
      2.8000           & 1.01                      & 11                                 & $\pm$                 & 4                 & 1.45       & 1.04                                      & 0.31                                     & 49.8  & $\pm$ & 17.6 & $\pm$ & 2.0  & $\pm$ & 2.2  & 51.6  & $\pm$ & 18.2 & $\pm$ & 2.0  & $\pm$ & 2.3  \\
      2.9000           & 105                       & 687                                & $\pm$                 & 37                & 1.45       & 1.03                                      & 0.30                                     & 29.3  & $\pm$ & 1.6  & $\pm$ & 1.2  & $\pm$ & 1.3  & 30.3  & $\pm$ & 1.7  & $\pm$ & 1.3  & $\pm$ & 1.4  \\
      2.9500           & 15.9                      & 114                                & $\pm$                 & 14                & 1.47       & 1.03                                      & 0.30                                     & 32.7  & $\pm$ & 4.1  & $\pm$ & 2.0  & $\pm$ & 1.5  & 33.6  & $\pm$ & 4.3  & $\pm$ & 2.1  & $\pm$ & 1.5  \\
      2.9810           & 16.1                      & 72                                 & $\pm$                 & 15                & 1.48       & 1.02                                      & 0.30                                     & 20.4  & $\pm$ & 4.3  & $\pm$ & 1.2  & $\pm$ & 0.9  & 20.9  & $\pm$ & 4.4  & $\pm$ & 1.2  & $\pm$ & 0.9  \\
      3.0000           & 15.9                      & 74                                 & $\pm$                 & 13                & 1.49       & 1.02                                      & 0.29                                     & 21.8  & $\pm$ & 3.9  & $\pm$ & 1.6  & $\pm$ & 1.0  & 22.3  & $\pm$ & 4.0  & $\pm$ & 1.6  & $\pm$ & 1.0  \\
      3.0200           & 17.3                      & 78                                 & $\pm$                 & 12                & 1.49       & 1.01                                      & 0.29                                     & 21.1  & $\pm$ & 3.3  & $\pm$ & 1.3  & $\pm$ & 0.9  & 21.4  & $\pm$ & 3.3  & $\pm$ & 1.3  & $\pm$ & 1.0  \\
      3.0800           & 126                       & 576                                & $\pm$                 & 34                & 1.59       & 0.92                                      & 0.27                                     & 23.5  & $\pm$ & 1.4  & $\pm$ & 1.0  & $\pm$ & 1.1  & 21.5  & $\pm$ & 1.3  & $\pm$ & 0.9  & $\pm$ & 1.0  \\
      \hline
      \hline
    \end{tabular}
  }
\end{table*}
}
\section{\boldmath Systematic Uncertainty}
{
Systematic uncertainties in the cross section measurement come from the luminosity measurement, tracking efficiency, PID efficiency, kinematic fit, signal and background shape, fitting range, radiative correction, MC sample size and the branching fraction of the decay $\phi \to K^{+}K^{-}$.

\begin{enumerate}
  \item The integrated luminosity is measured using large angle Bhabha events, with an uncertainty of $1.0\%$~\cite{data}.

  \item The tracking efficiency uncertainty is estimated to be $1.0\%$ for each track~\cite{phiKK}. Thus, $3.0\%$ is taken as the systematic uncertainty for the two pion and one kaon tracks.

  \item The PID efficiency uncertainty is estimated to be $1.0\% $ per $\pi^{\pm}$ and $1.0\%$ per $K^{\pm}$~\cite{phiKK}. So $3.0\%$ is taken as the systematic uncertainty on {the} PID efficiency.

  \item The uncertainty in $\BR(\phi \to K^{+}K^{-})$ is taken from the PDG~\cite{PDG}.

  \item The uncertainty from the kinematic fit comes from the inconsistency between the data and MC simulation of the helix parameters. Following the procedure described in Ref.~\cite{kinematic}, the helix parameters for the charged tracks of MC samples are corrected to eliminate the inconsistency during uncertainty study. The agreement of $\chi^{2}$ distributions between data and the MC simulation is significantly improved. Half of the difference between the selection efficiencies with and without the helix parameter correction is taken as the systematic uncertainty.

  \item Uncertainties due to the choice of signal shape, background shape and fitting range are estimated by introducing the changes below. The $\phi$ signal is described by a P-wave BW function convolved with a Gaussian function. To estimate the signal shape uncertainty, the signal shape is changed to the shape from the signal MC simulation convolved with a Gaussian function and the resulting difference is taken as the uncertainty from the signal model.  To estimate the background model uncertainty the background function is modified from a reversed ARGUS function to the function of $ f(M) = (M - M_{a})^{c} (M_{b} - M)^{d}$, where $M_{a}$ and $M_{b}$ are the lower and upper edges of the mass distribution while $c$ and $d$ are the parameters which were determined in the fit. The fit range  is extended from [0.98, 1.08]~GeV/$c^{2}$ to [0.98, 1.10]~GeV/$c^{2}$ to estimate the fit-range uncertainty. The differences between the number of signal events before and after the changes are taken as the systematic uncertainties.

  \item The uncertainty due to $\phi$ shape is examined by using an alternative fit involving the interferences of $\phi$ with $\omega$, $\rho$ and their excited states. The change of the fitted signal yield, 0.3\%, is assigned as a systematic uncertainty.

  \item Uncertainties in the possible distortions of the cross section line-shape introduce systematic uncertainties in the radiative correction factor and the efficiency. These are estimated by using the cross section line-shape function $\sigma = \sigma(\sqrt{s};p_{1},p_{2},...)$ obtained from the iteration described in Sec.~\ref{section:crosssection}, where $p_{i}(i=1,2,...)$ are the parameters which are determined in the fit. All parameters are randomly varied within their uncertainties and the resulting parametrization of the line-shape is used to recalculate (1 + $\delta$), $\epsilon$ and the corresponding cross sections. This procedure is repeated 1000 times and the standard deviation of the resulting cross sections is taken as a systematic uncertainty.

  \item The uncertainty from the MC sample size is estimated by the number of generated events.

\end{enumerate}

The first four sources of uncertainty are correlated between different energies, which give a total 4.4\% contribution to each energies. Other systematic uncertainties are uncorrelated.
\section{\boldmath Line-shape fitting}
{
In the cross section lineshape of the \process process, clear structure can be seen around 2.1~GeV, which is identified as the $\phi(2170)$ resonance. To parameterize its mass and width, the cross sections are fitted by a Breit-Wigner function represent $\phi(2170)$ and a continuum contribution:
\begin{equation}
  \begin{aligned}
     & \sigma=|f_{r}(\sqrt{s})\phi_{\rm P}+f_{c}(\sqrt{s})|^{2},                                                                                                                    \\
     & f_{r}(\sqrt{s}) = \frac{M_{r}}{\sqrt{s}} \frac{\sqrt{12\pi\varGamma_{ee}Br\varGamma_{r}}}{s-M^{2}+iM_{r}\varGamma(\sqrt{s})}\sqrt{\frac{\varPhi(\sqrt{s})}{\varPhi(M_{r})}}, \\
     & f_{c}(\sqrt{s}) = (p_{0} + p_{1}\cdot\sqrt{s} + p_{2}\cdot s )\cdot\sqrt{\varPhi(\sqrt{s})} ,                                                                                \\
  \end{aligned}
\end{equation}
where $\sigma$ represents the cross sections; $f_{r}$ and $f_{c}$ represent the contributions from resonance and continuum shapes; $M_{r}$ and $\varGamma_{r}$ are the mass and width of the resonant structure; $\varGamma_{ee}Br$ is the electric partial width times the branching fraction of the resonance decaying to corresponding intermediate states; $\varPhi$ is the phase space factor of the process \process calculated by using the method in Ref.~\cite{renew}; $\varGamma(\sqrt{s})=\varGamma_{r}\cdot(\varPhi(\sqrt{s})/\varPhi(M_{r}))$ is energy-dependent width; $\phi_{\rm P}$ is the phase angle between two components and $p_{0}$, $p_{1}$ and $p_{2}$ represent free parameters for the continuum shape.

In the fit of cross section lineshapes, the minimized $\chi^{2}$ constructed by incorporating both statistical and systematical uncertainties, as described in Ref.~\cite{covariance}, after considering the correlated and uncorrelated terms as formula
\begin{equation}
  \begin{aligned}
    \chi^2=\varDelta X^TM^{-1}\varDelta X,
  \end{aligned}
\end{equation}
where $\varDelta X$ is the difference between the measured cross section and the expected value calculated by function at each c.m. energy. The $M$ is the covariance matrix of elements
\begin{equation}
  \begin{aligned}
    M_{ij}=\begin{cases}
             \left( \varDelta _{i}^{\rm sys} \right) ^2+\left( \varDelta _{\mathrm{i}}^{\rm sta} \right) ^2, i=j        \\
             \left( \sigma _i\cdot \varepsilon _s \right) \cdot \left( \sigma _j\cdot \varepsilon _s \right) ,   i\ne j \\
           \end{cases},
  \end{aligned}
\end{equation}
where the index $i(j)$ represents the $i(j)$-th data set; the $\varDelta^{\rm sta}_{i}$ is the asymmetrically statistical uncertainty for $i$-th data set; the $\varDelta^{\rm sys}_{i}$ is the total systematic uncertainty and the $\varepsilon_{s}$ is the relative correlated systematic uncertainty.

The fit results are shown in Fig.~\ref{figure:fit} and Table~\ref{table:fit}.

\begin{figure*}[!htbp]
  \begin{center}
    \begin{overpic}[width=0.32\textwidth]{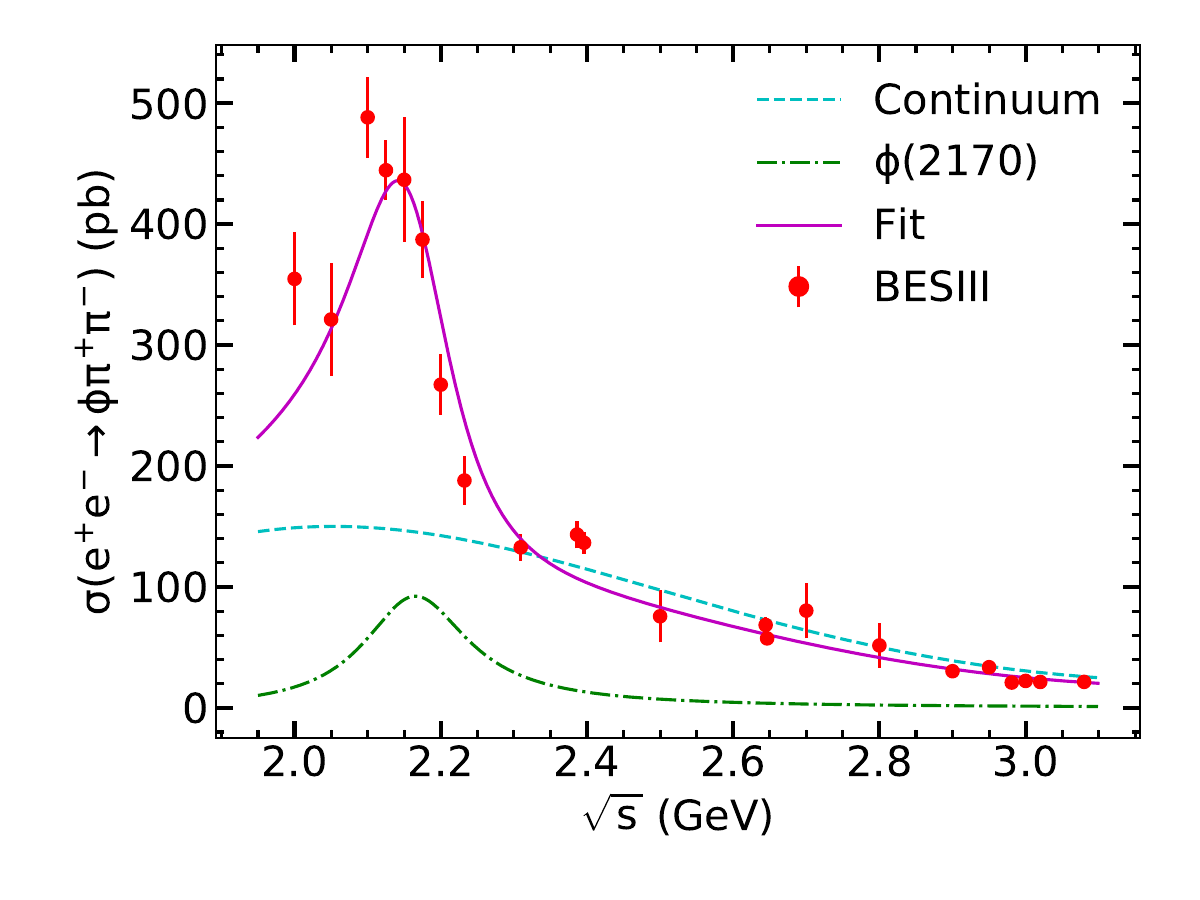}
      \put(80,30){\small\bfseries{(a)}}
    \end{overpic}
    \begin{overpic}[width=0.32\textwidth]{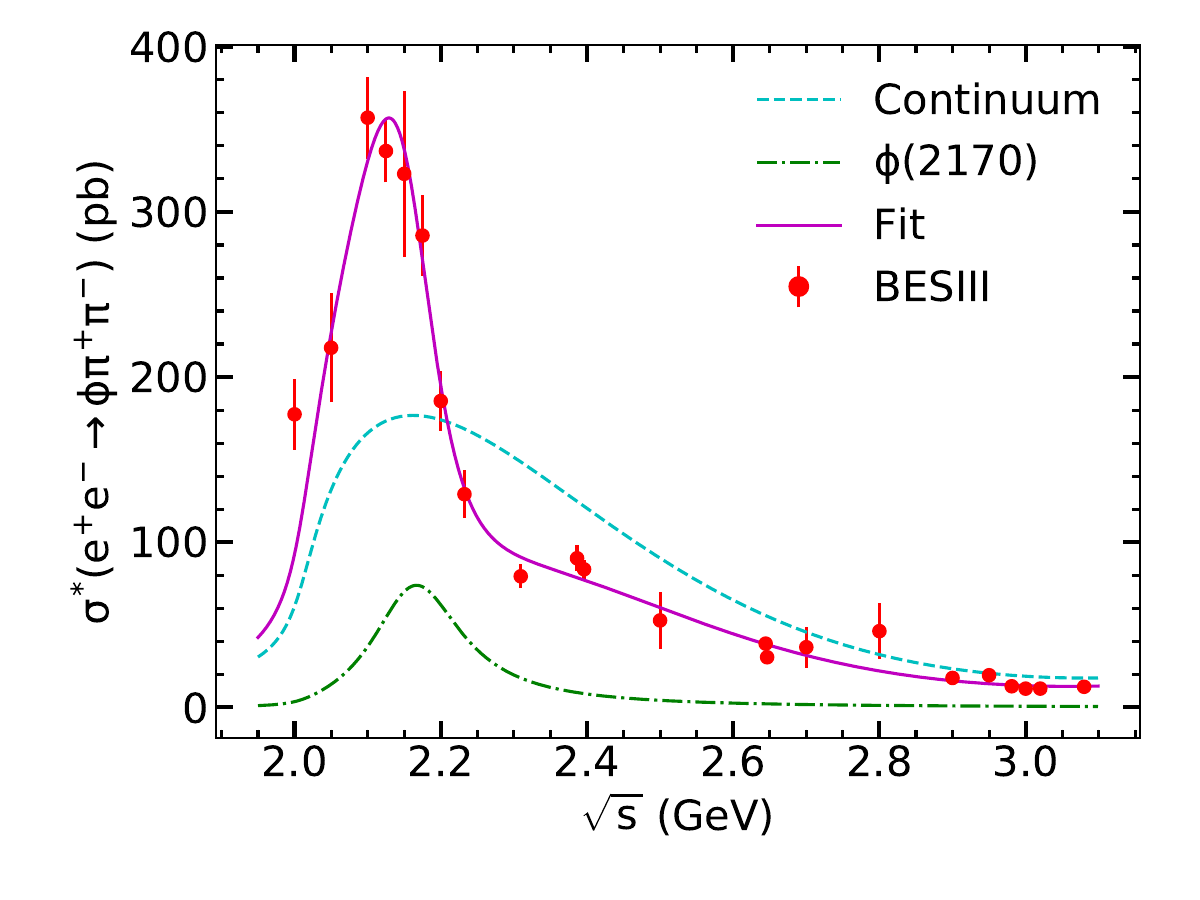}
      \put(80,30){\small\bfseries{(b)}}
    \end{overpic}
    \begin{overpic}[width=0.32\textwidth]{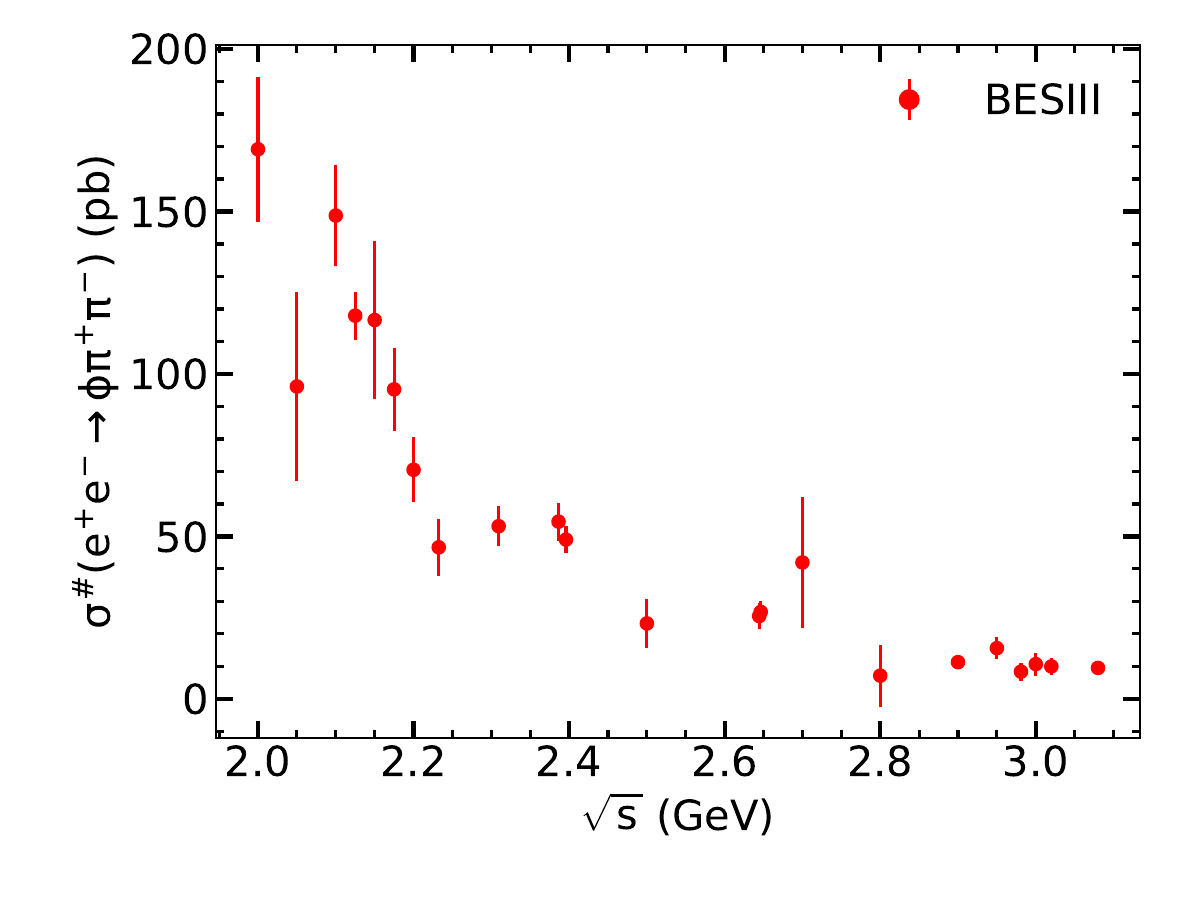}
      \put(80,30){\small\bfseries{(c)}}
    \end{overpic}
    \caption{The fit results of the cross sections of the process \process with events in (a) full $M_{\pi^+\pi^-}$ range, (b) $M_{\pip\pim}\in [0.85, 1.1]\unitmgev$, and (c) $M_{\pip\pim}\notin [0.85, 1.1]\unitmgev$ GeV$/c^2$. The red dots with error bars represent the obtained cross sections and the uncertainties incorporate statistical and uncorrelated systematic uncertainties. In (a) and (b), the purple solid line represnets the fit function and the green dotted line represents the contribution from $\phi(2170)$. The fits in (a) and (b) show the constructive interference results within two parallel solutions.}
    \label{figure:fit}
  \end{center}
\end{figure*}

To further explore the properties of $\phi(2170)$, we examine the cross section lineshape with events within $M_{\pip\pim}\in [0.85, 1.1]\unitmgev$, with the contribution of the $\ee\to\phi f_{0}(980)$ subprocess enhanced. The adjusted cross sections ($\sigma^{*}$) are fitted with the same method, with the phase space factors are replaced by those of the process $\ee\to\phi f_{0}(980)$~\cite{renew}. The results of the fit to this cross section lineshape are shown in Fig.~\ref{figure:fit}(b) and Table~\ref{table:fit}.

\begin{table}[!ht]
  \centering
  \caption{Results of the fits to the cross sections of \process within full $M_{\pi^+\pi^-}$ range and within $M_{\pip\pim}\in [0.85, 1.1]\unitmgev$, where mass is reported in$\unitmmev$; width is reported in$\unitemev$; $\phi_{\rm P}^{\rm D}$ and $\phi_{\rm P}^{\rm C}$ are phase angles in radian between two components with destructive and constructive interference, respectively.}
  \label{table:fit}
  \normalsize
  \begin{tabular}{c p{2.5cm}<{\centering} p{2.5cm}<{\centering}}
    \toprule
    Parameter                 & $\sigma$       & $\sigma^{*}$   \\
    \hline
    $M_r(\phi(2170))$         & $2171\pm12$    & $2178\pm20$    \\
    $\varGamma_r(\phi(2170))$ & $115\pm28$     & $140\pm36$     \\
    $\phi_{\rm P}^{\rm D}$    & $1.01\pm0.14$  & $1.13\pm0.06$  \\
    $\phi_{\rm P}^{\rm C}$    & $-1.87\pm0.16$ & $-1.71\pm0.12$ \\
    $\chi^{2}/{\rm ndf}$      & $28/15$        & $23/15$        \\
    \hline
    \hline
  \end{tabular}
\end{table}

In both fitting procedures, two multi-solutions are found with equal fit quality ($\chi^{2}/{\rm ndf}$). The figures above only show one of them. The statistical significance of $\phi(2170)$ is greater than $10\sigma$ for each solution. The uncertainties associated with the fit procedure include effects from the choice of continuum shape. Since the $\phi f_{0}(980)$ enhanced cross sections are closer to the nature of $\phi(2170)$, the fit result with $M_{\pip\pim}\in [0.85, 1.1]\unitmgev$ is taken as the nominal result. To estimate the systematic uncertainty, an alternative fit is carried out by using an exponential function to describe continuum shape. The difference between the nominal and alternative fit results are considered as the systematic uncertainty. The systematic uncertainties of the $\phi(2170)$ resonance mass and width are obtained to be $5\unitmmev$ and $16\unitemev$.

For these two fits, we have also tried to fit the cross section lineshapes by adding R(2400) in the fit. However, the statistical significance of R(2400) is no more than $2\sigma$.

In addition, we examined the cross section lineshape of $e^+e^-\to \phi\pi^+\pi^-$ with events in $M_{\pip\pim} \notin [0.85, 1.1]\unitmgev$, as shown in Fig.~\ref{figure:fit}. A similar fit is also performed on this lineshape, but no resonance structure with statistical significance greater than $3\sigma$ is found.
}
\section{\boldmath Conclusion}
In summary, the cross sections of the \process process are measured using data samples collected with the BESIII detector at 22 center-of-mass energies from 2.00 GeV to 3.08 GeV. The measured cross section is consistent with previous results from the \babar~\cite{renew}, Belle~\cite{confirmed:2} and BESIII~\cite{LiuF} experiments, but with improved precision.

In the cross section lineshapes of $e^+e^-\to\phi\pi^+\pi^-$ in full $M_{\pi^+\pi^-}$ range and in \mbox{$M_{\pi^+\pi^-}\notin [0.85,1.1]~{\rm GeV}/c^2$}, the $\phi(2170)$ resonance is clearly observed. For the last case, its mass and width are determined to be \mbox{$M=2178~\pm~20~\pm~5\unitmmev$} and \mbox{$\varGamma=140~\pm~36~\pm~16\unitemev$}, respectively, where the first uncertainties are statistical and the second ones are systematic. The central value of the $\phi(2170)$ width obtained in this work is consistent with existing results~\cite{confirmed:1, confirmed:2, renew, update:1, update:2}. However, no significant $\phi(2170)$ is observed in the cross section lineshape of $e^+e^-\to \phi \pi^+\pi^-$ with \mbox{$M_{\pi^+\pi^-}\notin [0.85,1.1]~{\rm GeV}/c^2$}.

In addition, no clear structure around 2.4~GeV has been found in this analysis. Since this structure at the same energy has been seen in the $K^{+}K^{-}f_{0}(980)$ mode with $f_{0}(980) \to \pi^{+} \pi^{-}$ and $\pi^{0} \pi^{0}$~\cite{observed1, observed2}, a future study of this channel with an amplitude analysis will be helpful to improve knowledge of the R(2400) state.

\acknowledgements
The BESIII collaboration thanks the staff of BEPCII, the IHEP computing center and the super computing center of USTC for their strong support. This work is supported in part by National Key Basic Research Program of China under Contracts Nos. 2020YFA0406400, 2020YFA0406300; National Natural Science Foundation of China (NSFC) under Contracts Nos. 11625523, 11635010, 11735014, 11822506, 11835012, 11935015, 11935016, 11935018, 11961141012, 12022510, 12025502, 12035009, 12035013, 12061131003, 11705192, 11950410506, 12061131003; the Chinese Academy of Sciences (CAS) Large-Scale Scientific Facility Program; Joint Large-Scale Scientific Facility Funds of the NSFC and CAS under Contracts Nos. U1732263, U1832207, U1832103, U2032111; CAS Key Research Program of Frontier Sciences under Contract No. QYZDJ-SSW-SLH040; 100 Talents Program of CAS; INPAC and Shanghai Key Laboratory for Particle Physics and Cosmology; ERC under Contract No. 758462; European Union Horizon 2020 research and innovation programme under Contract No. Marie Sklodowska-Curie grant agreement No 894790; German Research Foundation DFG under Contracts Nos. 443159800, Collaborative Research Center CRC 1044, FOR 2359, FOR 2359, GRK 214; Istituto Nazionale di Fisica Nucleare, Italy; Ministry of Development of Turkey under Contract No. DPT2006K-120470; National Science and Technology fund; Olle Engkvist Foundation under Contract No. 200-0605; STFC (United Kingdom); The Knut and Alice Wallenberg Foundation (Sweden) under Contract No. 2016.0157; The Royal Society, UK under Contracts Nos. DH140054, DH160214; The Swedish Research Council; U. S. Department of Energy under Contracts Nos. DE-FG02-05ER41374, DE-SC-0012069

\bibliographystyle{apsrev4-1}
\bibliography{main.bib}
\end{document}